\documentclass[10pt,conference]{IEEEtran}
\IEEEoverridecommandlockouts
\usepackage{cite}
\usepackage{amsmath,amssymb,amsfonts}
\usepackage{algorithmic}
\usepackage{graphicx}
\usepackage{textcomp}
\usepackage{xcolor}
\usepackage{subcaption}
\usepackage[left=1.2cm,right=1.62cm,top=1.9cm, bottom=4.16cm]{geometry}

\usepackage[nolist,nohyperlinks]{acronym}

\def\BibTeX{{\rm B\kern-.05em{\sc i\kern-.025em b}\kern-.08em
    T\kern-.1667em\lower.7ex\hbox{E}\kern-.125emX}}
\IEEEoverridecommandlockouts\IEEEpubid{\makebox[\columnwidth]{ 978-1-6654-3540-6/23/\$31.00 \copyright 2023 IEEE \hfill} \hspace{\columnsep}\makebox[\columnwidth]{ }}  
\begin{document}
\begin{acronym}
\acro{HAR}{High Aspect Ratio}
\acro{LHCP}{Left-Hand Circular Polarization}
\acro{SLL}{Side Lobe Level}
\acro{GLL}{Grating Lobe Level}
\acro{FoV}{Field of View}
\end{acronym}

\title{Satellite Swarms for Narrow Beamwidth Applications\\

\thanks{This work was supported by the Luxembourg National Research Fund (FNR), through the CORE Project (ARMMONY): Ground-based distributed beamforming harmonization for the integration of satellite and Terrestrial networks, under Grant FNR16352790.}
}

\author{\IEEEauthorblockN{1\textsuperscript{st} \small{Juan A. V\'asquez-Peralvo}}
\IEEEauthorblockA{\small{\centering{\textit{Interdisciplinary Centre }}}\\
\small{\centering{\textit{for Security, Reliability,}}}\\
\small{\centering{\textit{and Trust (SnT)}}}\\
\small{\centering{\textit{Universit\'e du Luxembourg}}}\\
\small{\centering{Luxembourg, Luxembourg}} \\
\small{\centering{orcid.org/0000-0001-7304-095X}}}
\and
\IEEEauthorblockN{2\textsuperscript{rd} \small{Juan Carlos Merlano Duncan}}
\IEEEauthorblockA{\small{\centering{\textit{Interdisciplinary Centre }}}\\
\small{\centering{\textit{for Security, Reliability,}}}\\
\small{\centering{\textit{and Trust (SnT)}}}\\
\small{\centering{\textit{Universit\'e du Luxembourg}}}\\
\small{\centering{Luxembourg, Luxembourg}} \\
\small{orcid.org/0000-0002-9652-679X}} \\\and
\IEEEauthorblockN{3\textsuperscript{th} \small{Geoffrey Eappen}}
\IEEEauthorblockA{\small{\centering{\textit{Interdisciplinary Centre }}}\\
\small{\centering{\textit{for Security, Reliability,}}}\\
\small{\centering{\textit{and Trust (SnT)}}}\\
\small{\centering{\textit{Universit\'e du Luxembourg}}}\\
\small{\centering{Luxembourg, Luxembourg}} \\
\small{orcid.org/0000-0002-4065-3626}}
\and
\IEEEauthorblockN{4\textsuperscript{nd} \small{Symeon Chatzinotas}}
\IEEEauthorblockA{\small{\centering{\textit{Interdisciplinary Centre }}}\\
\small{\centering{\textit{for Security, Reliability,}}}\\
\small{\centering{\textit{and Trust (SnT)}}}\\
\small{\centering{\textit{Universit\'e du Luxembourg}}}\\
\small{\centering{Luxembourg, Luxembourg}} \\
\small{orcid.org/0000-0001-5122-0001}}
\and
}

\maketitle

\begin{abstract}
Satellite swarms have recently gained attention in the space industry due to their ability to provide extremely narrow beamwidths at a lower cost than single satellite systems.
This paper proposes a concept for a satellite swarm using a distributed subarray configuration based on a 2D normal probability distribution. The swarm comprises multiple small satellites acting as subarrays of a big aperture array limited by a radius of 20000$\lambda_0$ working at a central frequency of 19 GHz. The main advantage of this approach is that the distributed subarrays can provide extremely directive beams and beamforming capabilities that are not possible using a conventional antenna and satellite design. The proposed swarm concept is analyzed, and the simulation results show that the radiation pattern achieves a beamwidth as narrow as 0.0015$^\circ$ with a maximum side lobe level of 18.8 dB and a grating lobe level of 14.8 dB. This concept can be used for high data rates applications or emergency systems.
\end{abstract}

\begin{IEEEkeywords}
satellite swarm, direct radiating arrays, beamforming
\end{IEEEkeywords}

\section{Introduction}
Satellite swarms are constellations of low-cost small-size satellites working together to accomplish specific tasks, for instance, planetary observation \cite{borgue2022developing}, magnetospheric studies \cite{rajan2020apis}, space-to-earth communication \cite{budianu2015swarm}, to mention a few. Compared with a conventional one-satellite for-service scheme, swarm satellites have multiple advantages, e.g., improved global coverage, low costs, narrow beamwidths, redundancy, and high data rates. Nevertheless, there are some challenges in the design of swarm satellites, for instance, synchronization, orbital debris generated by the swarm, and inter-satellite communications \cite{farrag2021satellite}.

Some authors have proposed different approaches to obtain beams with a very narrow beamwidth in the literature. For instance, in the first approach presented in \cite{murao1998proposal}, authors have presented an array of 100 elements using 10 meters circular and hexagonal aperture arrays to obtain directive beams with a minimum beamwidth of 0.08$^\circ$. In the second approach, authors use spacecraft formation flying and reconfiguration control to assemble a high-dimension antenna in space to accomplish high-gain beams \cite{she2019constructing}. The third approach uses a similar concept using \ac{HAR}, where the authors propose different approaches, for instance, using deployable flat sub-antenna modules \cite{ASTScienceBluewalker3}, or independent mini disk satellites \cite{SpaceNewsDisksatProposal} to manufacturing very thin satellites to place high-size antennas and solar panels. Finally, other authors use deployable antennas like membrane reflectors \cite{lichodziejewski2003inflatably}, inflatable reflectors \cite{cassapakis1995inflatable}, umbrella reflectors \cite{duan2020large}, and radial rib antennas \cite{semler2010design}.
This paper proposes the concept of an array of 256 satellites separated at random distances dictated by a 2D normal probability distribution with a standard deviation $\sigma = 40000\lambda_0$, mean $\mu = 0$, and limited to a radius $r = 20000 \lambda_0$.

\section{Design}
The satellite swarm design assumes an ideal scenario, which involves perfect synchronization, a low-error positioning system, and continuous communication for each satellite.
The design of the full satellite swarm has been done in three stages described below.
\begin{itemize}
    \item The first stage consists design of the radiating element used to generate the sub-array in each satellite. 
    \item The second stage involves the design of the sub-array that will be placed in each satellite.
    \item The last stage consists in the constellation array design that will generate the final radiation pattern.
\end{itemize}
A more detailed description and design steps are given for each itemized stage.
\subsection{Unit Cell design}
\begin{figure}[!htp]
     \centering
        \begin{subfigure}[b]{0.45\textwidth}
         \centering
         \includegraphics[width=\textwidth]{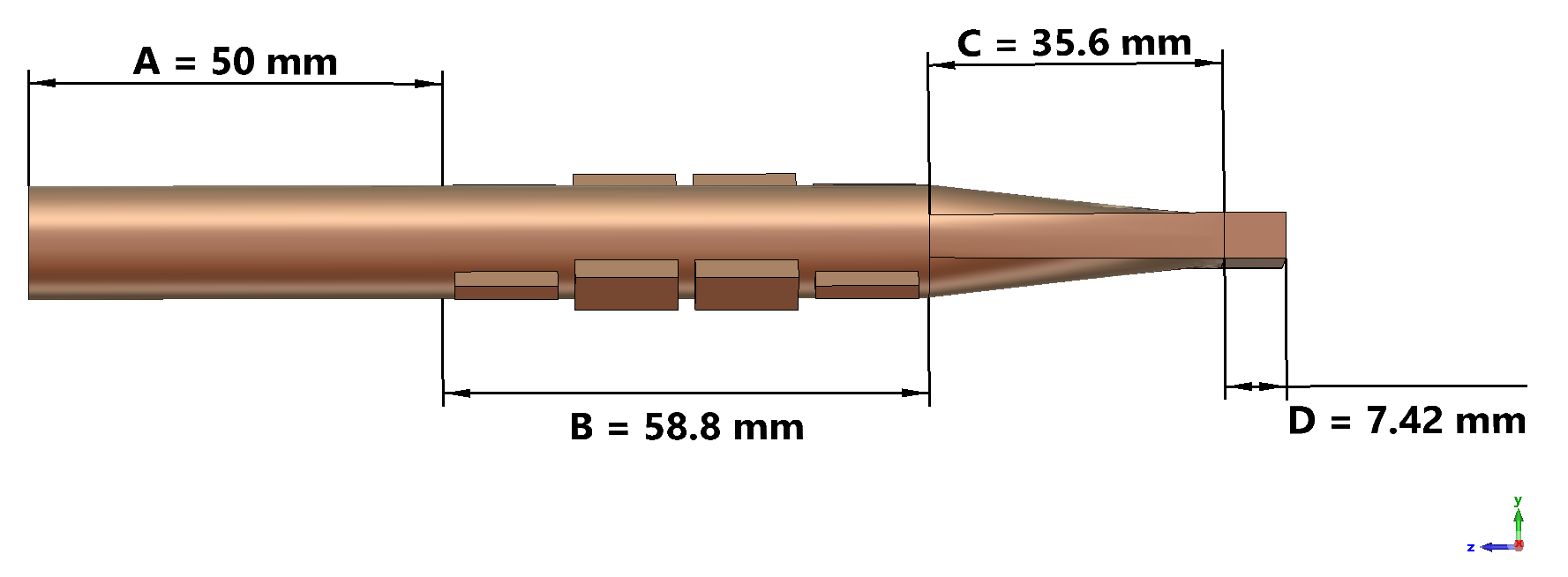}
         \caption{}
         \label{fig:Fig.Dimensionsa}
     \end{subfigure}
     \hfil
     \begin{subfigure}[b]{0.4\textwidth}
         \centering
         \includegraphics[width=\textwidth]{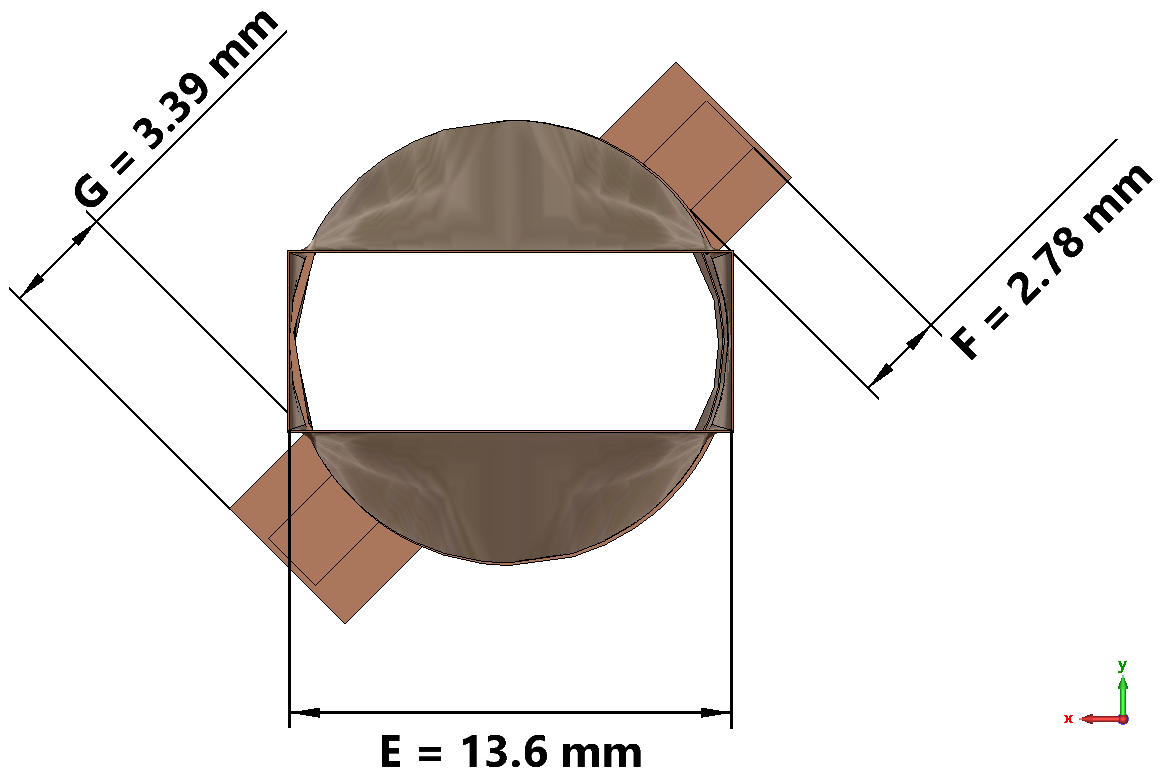}
         \caption{}
         \label{fig:Fig.Dimensionsb}
     \end{subfigure}
     \hfil
     \begin{subfigure}[b]{0.45\textwidth}
         \centering
         \includegraphics[trim=200 0 0 0, clip, width=\textwidth]{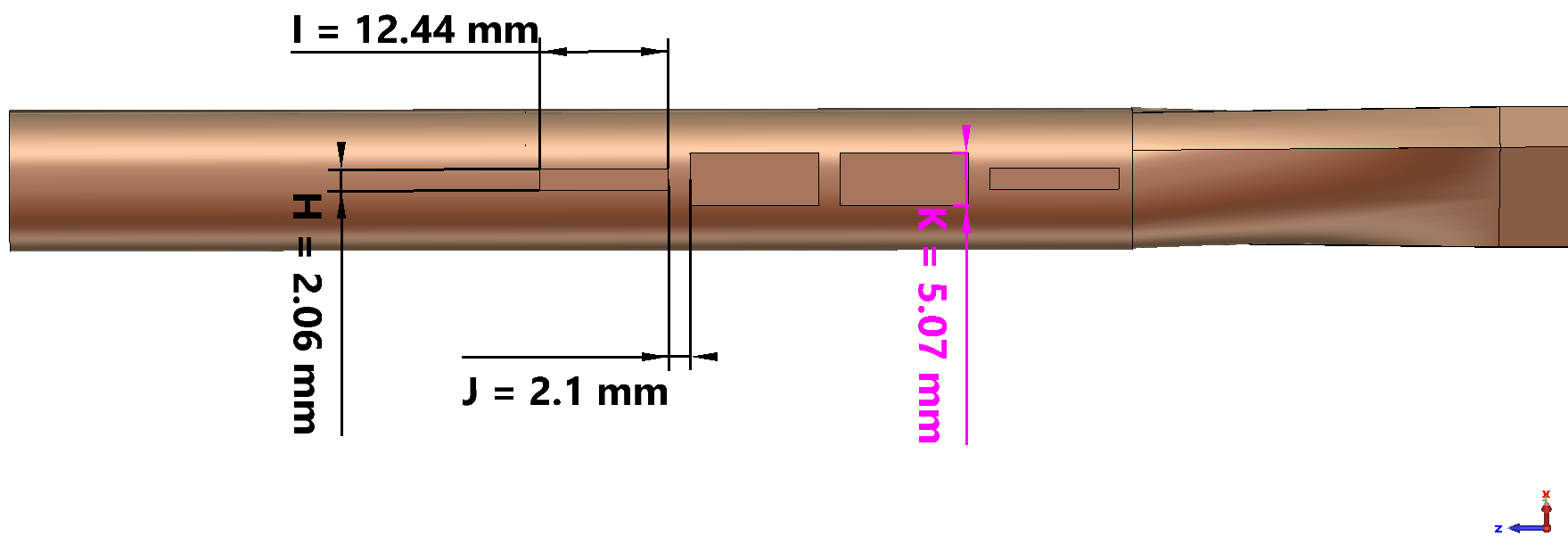}
         \caption{}
         \label{fig:Fig.Dimensionsc}
     \end{subfigure}
     
         \caption{Designed antenna dimensions. a) Lateral view. b) Front view. c) Lateral, rotated 45$^\circ$ view. }
        \label{fig:Fig.Dimensions}
\end{figure}

\begin{figure}[!htp]
     \centering
        \begin{subfigure}[b]{0.45\textwidth}
         \centering
         \includegraphics[width=\textwidth]{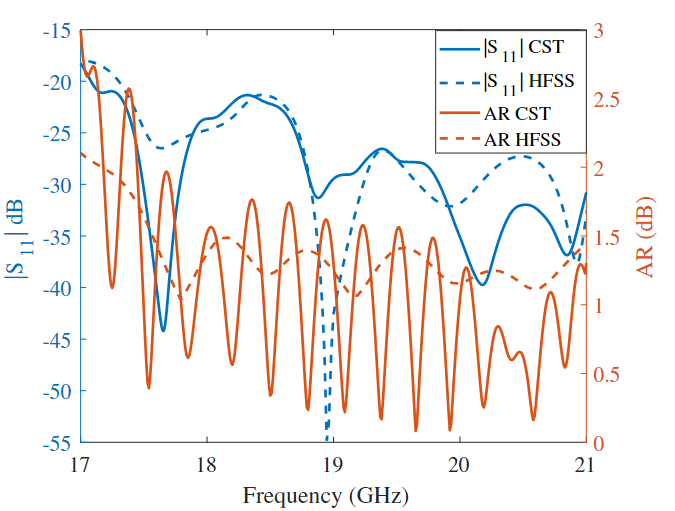}
         \caption{}
         \label{fig:Fig.S-Parameters}
     \end{subfigure}
     \hfil
     \begin{subfigure}[b]{0.45\textwidth}
         \centering
         \includegraphics[width=\textwidth]{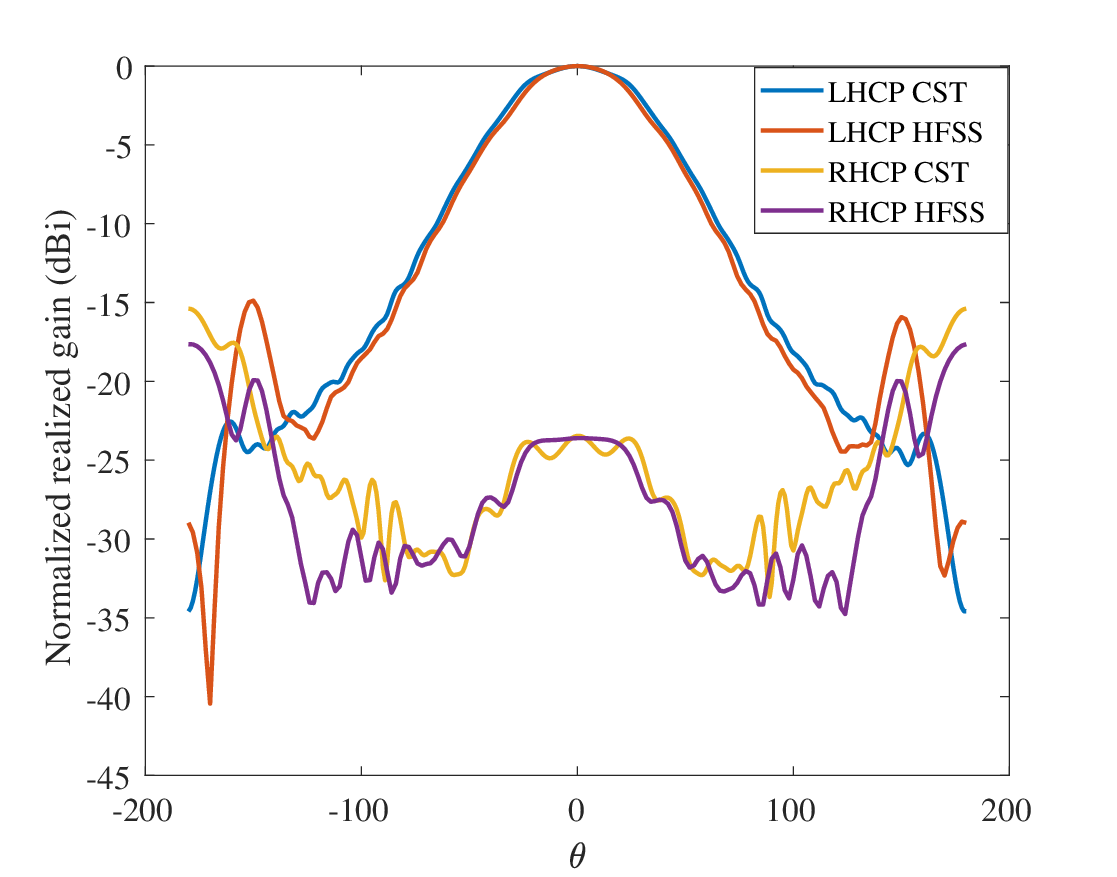}
         \caption{}
         \label{fig:Fig.RadPatter}
     
     \end{subfigure}
     
         \caption{Simulation results of the proposed circular polarized open-ended waveguide using CST Microwave Studio and HFSS. a) Reflection coefficient and axial ratio. b) Normalized radiation pattern.}
        \label{fig:Fig.Simulations} 
       
\end{figure}

The unit cell proposed in this design is an open-ended waveguide working in \ac{LHCP} to reduce the losses due to dielectrics as much as possible. A double-grooved located at 45$^\circ$ from the sagittal plane is used. Each double-grooved consists of two central and two edge grooves that will generate an offset of +/- 45$^\circ$ in the linear polarization, thus obtaining a \ac{LHCP}. The designed antenna is illustrated in Figure \ref{fig:Fig.Dimensions} and the full-wave simulation in Figure \ref{fig:Fig.Simulations}. The simulation results have been simulated using CST microwave studio, and for contrast, it has also been simulated in HFSS. The results show that the designed antenna will radiate efficiently within the bandwidth of interest, 17.7 - 21.2 GHz. Moreover, the radiation pattern is symmetric and with very low cross-polarization. Finally, the axial ratio in the bandwidth of interest is under -3 dB, which goes in hand with low cross-polarization results.

\section{Subarray design}
The sub-arrays design stage plays a crucial role in determining the key features of the final radiation pattern of an antenna array. One of the most significant factors to consider during this stage is the dimensioning of the subarray, as it heavily influences the scanning losses of the array and its overall gain.

For instance, the proposed sub-array, which comprises 422 elements distributed in a circular aperture with a triangular lattice of 0.857$\lambda_0$, has several advantages that help optimize the radiation pattern. The lattice and aperture configuration aid in achieving a low side lobe level (\ac{SLL}) and delaying the onset of grating lobes, both of which can significantly impact the final radiation pattern obtained in the array design.

With the aforementioned antenna parameters, we can achieve a scanning angle with a maximum loss of -3 dB $\alpha_{-3dB} = 3.3 ^\circ$. However, it is essential to note that there exists a trade-off between maximum gain and high scanning angles, both of which depend primarily on the number of elements in each sub-array.

\begin{figure}[!htp]
     \centering
        \begin{subfigure}[b]{0.45\textwidth}
         \centering
         \includegraphics[width=\textwidth]{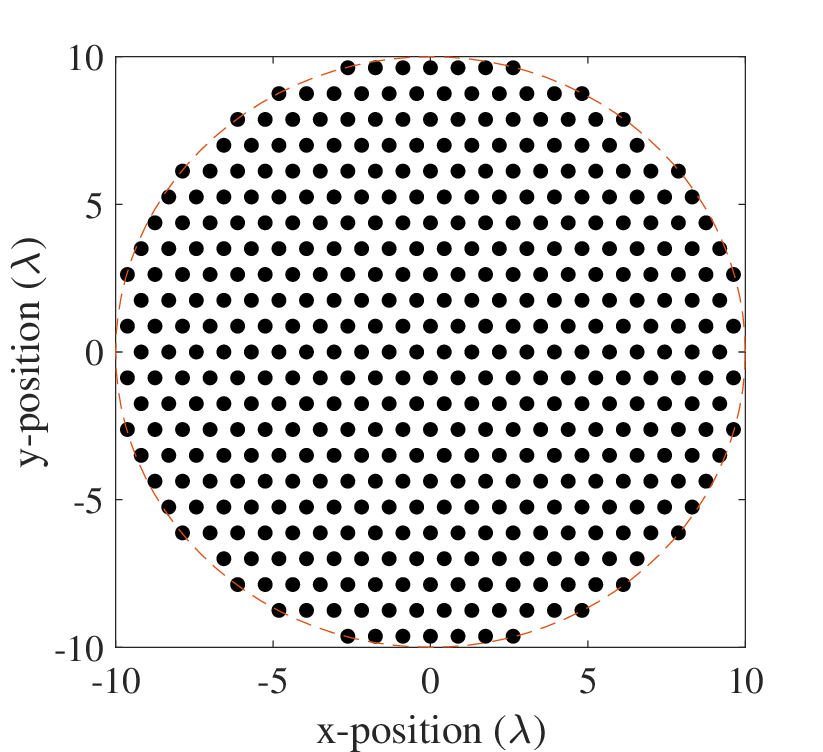}
         \caption{}
         \label{fig:Fig.SubArrayView}
     \end{subfigure}
     \hfil
     \begin{subfigure}[b]{0.45\textwidth}
         \centering
         \includegraphics[width=\textwidth]{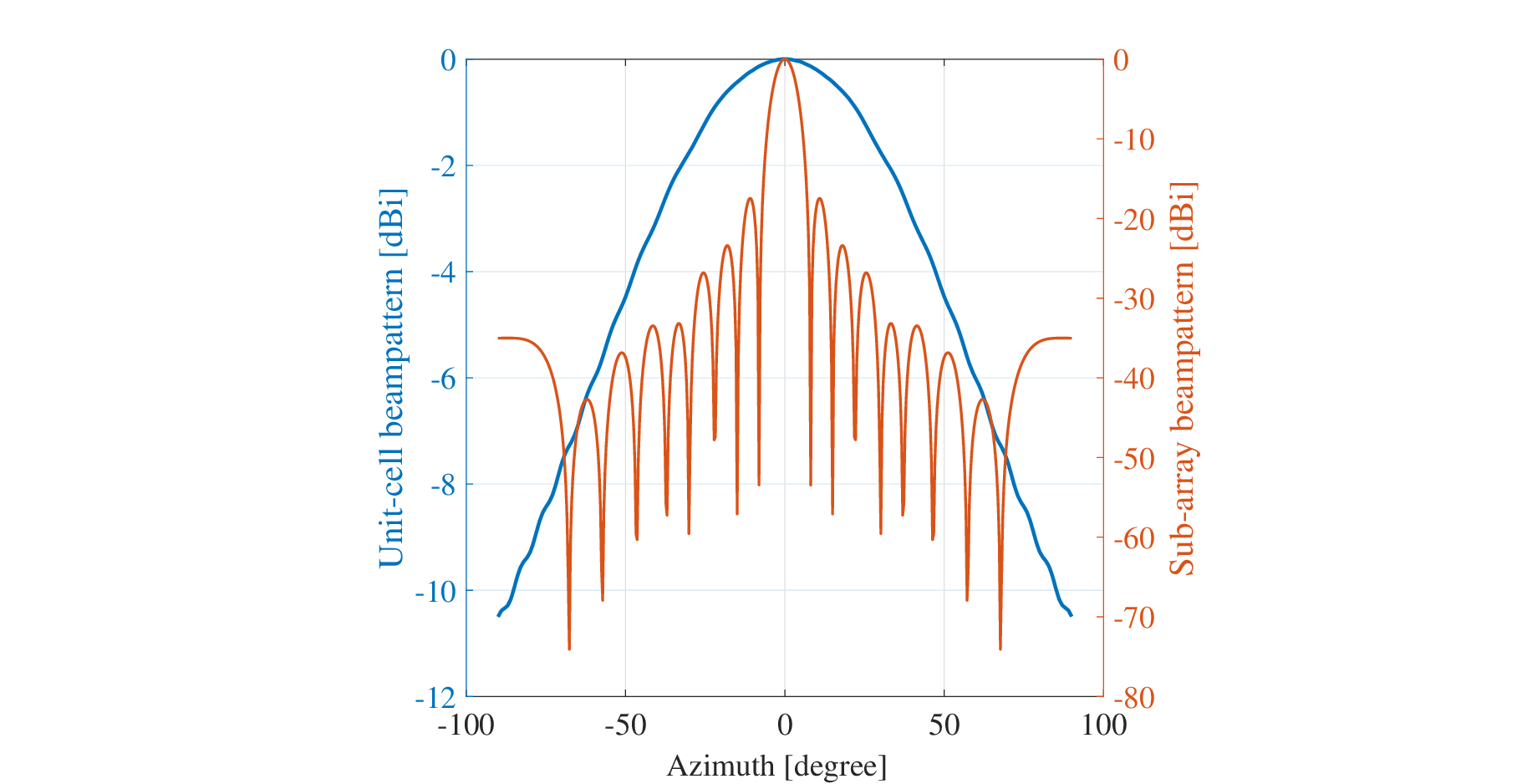}
         \caption{}
         \label{fig:Fig.RadaitionPatternComp}
     
     \end{subfigure}
     
         \caption{SubArray antenna design and simulation results. a) Sub-array element distribution. b) Comparison between unit cell and sub-array radiation pattern.}
        \label{fig:Fig.Subarray}
        
\end{figure}

The distribution of the antenna elements and the azimuth cut radiation pattern can be visualized in Figure \ref{fig:Fig.Subarray}, illustrating the effectiveness of the proposed subarray design in achieving a desirable radiation pattern. Overall, careful consideration of the subarray design is essential in optimizing the performance of an antenna array.

Considering the aperture and depth of the array, which is (10$\lambda_0$ by 9.61 $\lambda_0$), and the rest of the needed payload, the type of satellite required is a micro-satellite.

It is also worth noting that the proposed subarray design can be further improved and extended through various techniques, such as adaptive beamforming and advanced signal processing algorithms, which can help enhance the array's capabilities in terms of signal detection, localization, and tracking.

\section{Array design}

\begin{figure}[!htp]
     
         \centering
         \includegraphics[ width=0.9\columnwidth]{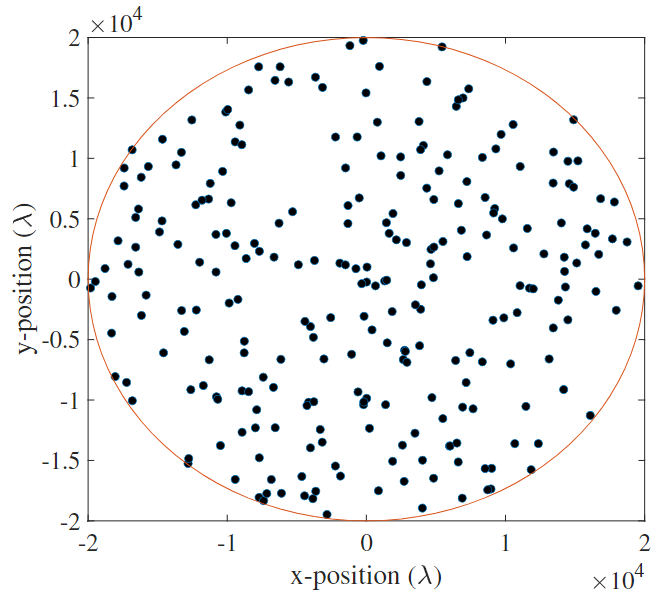}
         \caption{256 Satellite distribution in a maximum radius of 20000$\lambda_0$.}
         \label{fig:Fig.satelliteDistribution}

\end{figure}

The final array design has to have a big overall aperture; for this, each of the micro-satellites carrying each sub-array has to be separated at a reasonable distance, hundreds of wave-lengths, which simultaneously will avoid collisions. To overcome this challenge, in this paper, we have distributed the satellites using normal probability distribution with a mean of $\mu=0$ and a standard deviation of $\sigma=40000\lambda_0$. Considering this distribution and the sub-array has a directive beam with low \ac{SLL}, the final radiation pattern will have low \ac{SLL} and reduced onset of grating lobes. Moreover, to ensure no overlap between the satellites, a minimum distance separation of $d=500\lambda_0$ is set for each sub-array. A total of 256 satellites are then distributed in a radius of $r=20000\lambda_0$ to achieve the desired coverage area, as shown in Figure \ref{fig:Fig.satelliteDistribution}. However, it is important to note that the choice of the number of satellites, radius, and separation distance may vary depending on the specific application requirements. Finally, the formation of the satellites has to track this formation or a similar normal probability distribution to keep the required performance.
 \subsection{Simulation Results}
\begin{figure}[!htp]
     
         \centering
         \includegraphics[trim=0 0 0 0, clip, width=\columnwidth]{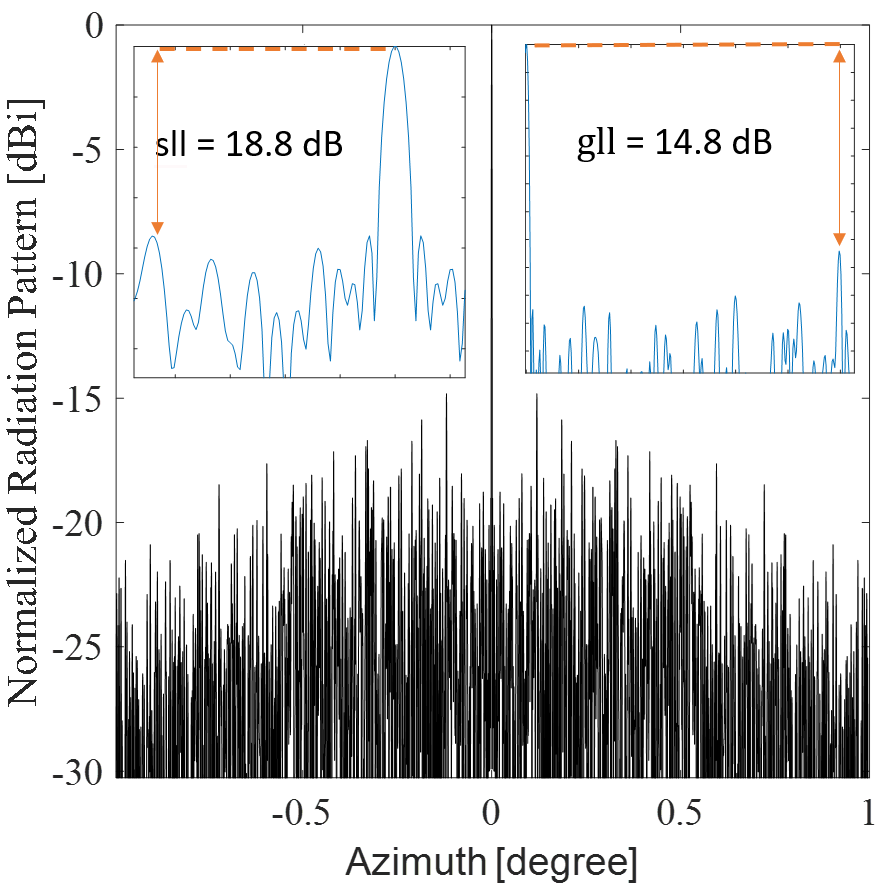}
         \caption{Array radiation pattern in the range of -1$^\circ$ to 1 $^\circ$. The left figure represents the highest nearby SLL, and the right figure represents the highest grating lobe level.}
         \label{fig:Fig.FinarResults}

\end{figure}

 Due to the very large size of the final array, the simulation has been carried out by programming the  array factor formula along with the previously calculated Sub-array radiation pattern in Matlab. Moreover, the extraction of different parameters like \ac{SLL}, beamwidth, and antenna footprint was also programmed from scratch in Matlab. To ensure accuracy, a granularity of 15000 points has been configured for the angle range from -1$^\circ$ to 1$^\circ$. The results obtained are presented in Figure \ref{fig:Fig.FinarResults}, which shows the cut at $\theta = 0^\circ$. To provide a more detailed analysis of the results, a zoomed-in view has been plotted, highlighting the main beam, having a $\theta_{-3dB} =0.0015^\circ$, and the nearest highest \ac{SLL} located at $\theta_{SLL}=$0.022$^\circ$, as well as the highest \ac{GLL} located at $\theta_{GLL}=$0.12$^\circ$. It is worth noting that the sub-array design has significantly reduced the \ac{SLL} in this case by 5 dB compared to the regular case, making it a remarkable improvement. Additionally, the highest \ac{GLL} of 14.8 $dB$ is comparable to far-located \ac{SLL}, which can be challenging to distinguish.

Assuming that the center of the satellite swarm is located at (0$^\circ$,50$^\circ$) in the geostationary orbit, we can estimate the coverage area and footprint of the array design. In particular, we need to ensure that the center of the beam is located in the city center of Luxembourg (49.612$^\circ$, 6.129$^\circ$).

\begin{figure}[!htp]
     
         \centering
         \includegraphics[trim=0 0 0 0, clip, width=\columnwidth]{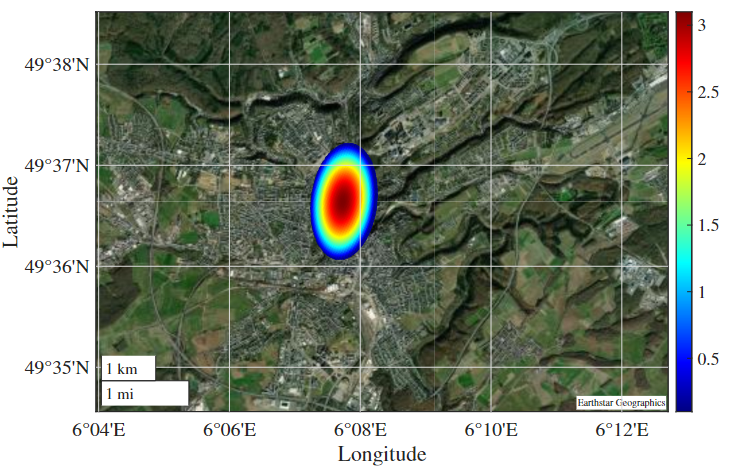}
         \caption{Half-power footprint pointing at the center of Luxembourg located at (49.612$^\circ$, 6.129$^\circ$). }
         \label{fig:Fig.FinarResultsFootprint}

\end{figure}

Based on these parameters, the footprint of the array will cover an area of approximately 1.7 $km^2$ within its half-power beamwidth, as shown in Figure \ref{fig:Fig.FinarResultsFootprint}. This is a crucial consideration for ensuring the array provides adequate coverage to the desired area. It highlights the importance of carefully selecting the location and orientation of the array design.

\section*{Acknowledgment}
This work was supported by the Luxembourg National Research Fund (FNR), through the CORE Project (ARMMONY): Ground-based distributed beamforming harmonization for the integration of satellite and Terrestrial networks, under Grant FNR16352790.
\section{Conclusions}
A normal distribution satellite swarm has been presented, analyzed, and simulated. The beamwidth obtained using the swarm can achieve a beamwidth $\theta_{-3dB}=$0.0015$^\circ$ with relatively low \ac{SLL} and \ac{GLL}. Moreover, considering the swarm placed in a geostationary position can cover an area of only 1.7 $km^2$, which makes it perfect for mobile communications, high-data-rate applications, and even for emergency communications, to mention a few examples. In addition to the presented work, other challenges have to be studied, for instance, the synchronization between swarm satellites, the application of FFT for beam steering, beamforming, interference control, and multi-beam scenarios application, among others. Furthermore, the beam can be further tailored by applying amplitude and phase control to, for instance, reduce \ac{SLL}, nulling in a particular direction, and changing the beamwidth. Finally, designing a satellite array with a narrow beamwidth requires careful consideration of several factors, including element spacing, sub-array separation, and the number and distribution of satellites. Considering these factors, it is possible to design a high-performance satellite array that meets the specific requirements of coverage area and \ac{FoV}, applied, for instance, to ultra-high speed satellite networks.
\bibliographystyle{IEEEtran}
\bibliography{bibliography.bib}

\end{document}